\documentclass[aps,pra,twocolumn,showpacs,superscriptaddress,floatfix]{revtex4}
\usepackage{amsmath}
\usepackage{amssymb}
\usepackage{graphicx}
\usepackage{color}
\usepackage{subfigure}
\usepackage{ulem}
\usepackage{enumitem} 
\usepackage{bm}

\definecolor{dgreen}{rgb}{0,0.5,0}
\definecolor{delete}{cmyk}{0.5,0,0,0}

\newcommand{\bra}[1]{\langle#1|}

\newcommand{\ket}[1]{|{#1}\rangle}

\newcommand{\Tr}{\mathop{\mathrm{Tr}}\nolimits}

\begin{document}
\title{Estimating Temperature via Sequential Measurements}
\author{Antonella De Pasquale}
\affiliation{NEST, Scuola Normale Superiore and Istituto Nanoscienze-CNR, Piazza dei Cavalieri 7, I-56126 Pisa, Italy}
\author{Kazuya Yuasa}
\affiliation{Department of Physics, Waseda University, Tokyo 169-8555, Japan}
\author{Vittorio Giovannetti}
\affiliation{NEST, Scuola Normale Superiore and Istituto Nanoscienze-CNR, Piazza dei Cavalieri 7, I-56126 Pisa, Italy}

\begin{abstract}
We study the efficiency of estimation procedures where the temperature of an external bath is indirectly recovered by monitoring the transformations induced on 
 a probing system that is put in thermal contact with the bath. In particular we compare 
the performances of sequential measurement schemes where the probe is initialized only once and measured repeatedly during its interaction with the bath, with those of
 measure \& re-prepare approaches where instead, after each interaction-and-measurement stage, the probe is reinitialized into the same fiduciary state. 
From our analysis it is revealed that the sequential approach, while being in general not capable of providing the best accuracy achievable, is nonetheless more versatile with respect
to the choice of the initial state of the probe, yielding on average smaller indetermination levels.
\end{abstract}
\maketitle

\section{Introduction}
Accurate temperature readings at nanoscales find applications in several research areas, spanning from materials science \cite{mat_sc1, mat_sc2, mat_sc3}, medicine and biology~\cite{bio1, bio2}, to quantum thermodynamics \cite{qtermo1,qtermo2,qtermo3}, where it is crucial for controlling the performances of quantum thermal devices. 
 The interest in this field is also motivated by the recent developments of nanoscale thermometry such as carbon nanothermometers~\cite{gao}, diamond sensors~\cite{diamond}, scanning thermal microscopes~\cite{stm}, \textit{etc}.
%%%%%%%%%%%%%
\begin{figure}[t]
\includegraphics[width=0.8\columnwidth]{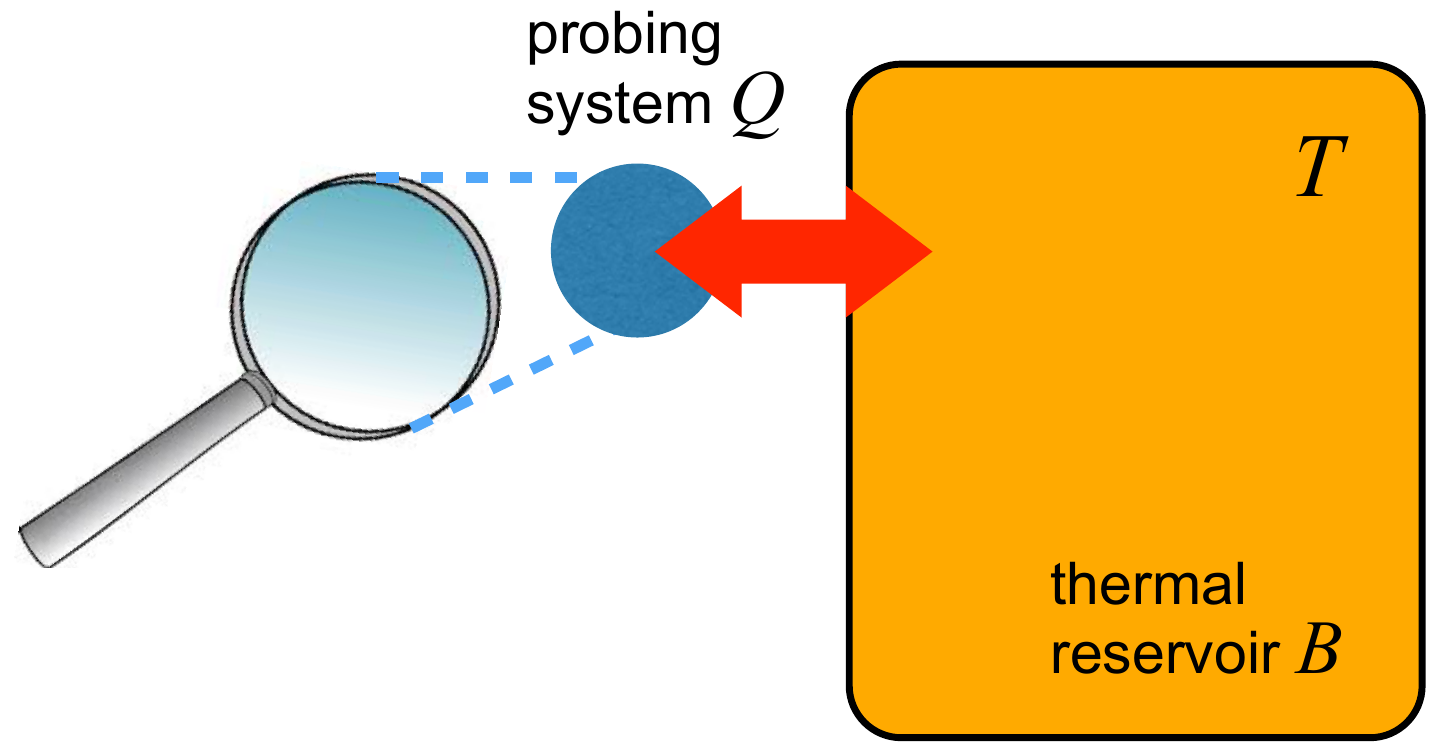} 
\caption{Schematic visualization of an indirect estimation of the temperature of a bath $B$ via the measurements on a probing system $Q$.}\label{figu1}
\end{figure}
%%%%%%%%%%%%%
Here we shall focus on a specific, yet rather general, thermometric task where the temperature of 
a sample characterized by a large number of subcomponents, also called reservoir,  is indirectly recovered by monitoring a small probe that is put in thermal contact with the reservoir: see Fig.~\ref{figu1}. 
Specifically the setting we consider is related to quantum thermometry, which aims to use low dimensional quantum systems (say qubits) as effective thermometers  to minimize the undesired disturbance on the sample: see e.g.~Refs.~\cite{correa1,betaloc,correa2} and references therein.
From a theoretical point view, the standard approaches to this kind of problems typically start from three hypotheses:
\begin{enumerate}[label=\textit{\roman*)}]
\item the reservoir is in a thermal state;
\item the probing system interacts for enough time with the bath so as to reach thermal equilibrium;
\item independent and identically distributed (IID)  measurements:  the experimentalist has at disposal a certain number of probes, prepared in the same input state, which interact with the bath and are measured independently. Equivalently the experimentalist might reinitialize the state of the single probe  after each measurement stage.
\end{enumerate}
Recently Correa \textit{et~al.}~\cite{correa1} proved that, under the above three assumptions, 
optimal thermometers correspond to employing 
atoms with a single energy gap and maximally degenerate first excited levels.  
On the other hand, if the interaction time with the reservoir is not long enough to allow complete thermalization of the probe (i.e.~hypothesis \textit{ii)} missed),  the maximal thermal sensitivity of the setup is reached by initializing the  probes in their ground states. Even more fundamental limitations emerge in the low temperature regime, in which the thermalization process might be prevented by the strong enough correlations between the probe and the sample~\cite{correa2}.

In this work we will concentrate on the drawbacks related to the IID assumption (hypothesis {\it iii)} of the list). In particular, the arbitrary initialization of independent probing systems at the beginning of the estimation procedure, or of a single probe at disposal after each measurement process, might encounter some obstructions, due to fundamental or practical reasons.  
One way to 
 circumvent such difficulty is to rely on sequential measurement schemes (SMSs)~\cite{guta, catana, kiilerich, burgarth},
 where  repeated consecutive measurements are performed on a single probe while
it is still in interaction  with the bath without reinitializing it. 
The performance of SMS will be therefore compared  with the IID protocol in different specific situations, taking the Fisher informations (FIs)~\cite{cramer,paris1,metrology1,paris2,metrology2} as the figure of merit for the corresponding temperature estimation accuracies. 
Quite interestingly we will find that, while in most cases optimality is attained by the IID approach, the SMS is more versatile as it is less affected by the choice of the initial state of the probe.  This phenomenon can be ascribed to the fact that in the SMS approach the probe is forced to gradually lose the memory of the initial condition, moving towards a fixed point configuration, which keeps track of the bath temperature.  Hence, the recursive character of SMS allows the probing system  to ``adapt'' to the reservoir, in such a way that even a non-optimal initialization of the probe can in the end provide a relatively good estimation of the temperature.

This paper is organized as follows. In Sec.~\ref{the_model}, after briefly describing the mathematical model for the Bosonic reservoir, we will present the IID and SMS strategies and show how to compute the FI in the two cases for generic measurements. A specific family of them will be selected in Sec.~\ref{sec:comp}, where we will also provide a numerical comparison between the two estimation schemes. Conclusions and final remarks are given in Sec.~\ref{sec:conclusions}.

\section{The model}\label{the_model}
Consider a Bosonic  thermal reservoir  $B$ of unknown temperature $T$, which we aim to recover by monitoring the relaxation dynamics induced on a probe $Q$, acting as a local thermometer in contact with $B$: see Fig.~\ref{figu1}. 
Since the detailed inner structure of $Q$ is expected to be irrelevant for discriminating between the performances of IID and SMS strategies, we take it to be a simple two-level system (qubit) and  describe the dynamical evolution it experiences when in contact with $B$ via a standard Markovian
Bloch master equation
 \cite{bosonic_res2,bosonic_res1}. Accordingly introducing the Pauli operators
 $\sigma_x$, $\sigma_y$, and $\sigma_z$, and the associated spin-flip operators 
 $\sigma_{\pm}=(\sigma_x \pm i \sigma_y)/2$, 
 the density matrix $\rho(t)$ of $Q$ will obey the following differential equation  \begin{equation}\label{LE}
\frac{d \rho (t)}{dt}= \mathcal{L}(\rho(t)),
\end{equation}
where $\mathcal{L}$ is the super-operator 
\begin{align}\label{eq:master} 
\mathcal{L}({\cdots})
= {}&{-\frac{i}{2}} \Omega [\sigma_z, ({\cdots})]_-   \nonumber \\
&{} +  \sum_{s= \pm1} \gamma_{s}    \left(
\sigma_{-s} ({\cdots}) \sigma_{s} -  \frac{1}{2}[ \sigma_{s}\sigma_{-s}, ({\cdots})]_+ \right)  ,
\end{align}
 with $[({\cdots}),({\cdots})]_\pm$ being the commutator ($-$) and anti-commutator ($+$) brackets. In this expression
 $\Omega$ is the characteristic frequency of $Q$, while   $\gamma_{+}$ and $\gamma_{-}$ are the relaxation constants associated to  the decay and excitation processes, respectively, given by
 \begin{equation}
 \gamma_{+} = (1+N_\mathrm{th}) \gamma, \qquad
 \gamma_{-}=N_\mathrm{th} \gamma, 
 \end{equation} 
where  $\gamma$ is a temperature independent parameter that gauges the   strength  of the $Q$-$B$ interactions, while $N_\mathrm{th}$ 
is the average thermal number of the Bosonic bath excitations, which are at resonance with the probe  and is responsible for imprinting the bath temperature $T$ into $\rho(t)$, i.e.~given by
\begin{equation} 
N_\mathrm{th}=\frac{1}{e^{\beta}-1}, \qquad   \beta = \hbar \Omega/({k_B T}),
\label{defN}
\end{equation} 
with $k_B$ being the Boltzmann constant.
Our goal is to determine the value of the temperature $T$
by monitoring the evolution of $Q$ induced by the thermalization process described by the master equation~(\ref{LE})~\cite{NOTA}. Adopting the 
positive operator-valued measure (POVM) formalism~\cite{NIELSEN}, we select a family of completely positive quantum maps $\{\mathcal{M}_s\}$
fulfilling the normalization condition $\sum_s \mathcal{M}_s = \mathcal{I}$, with $\mathcal{I}$ being the identity super-operator.
 When applied to a generic state $\rho$ of $Q$ this measurement provides the  outcome $s=\pm1$ with probability 
 \begin{equation} 
  P(s|\rho)=\Tr[\mathcal{M}_{s}(\rho) ] ,
  \label{EXPP10}
 \end{equation}
 while   inducing the following instantaneous quantum jump on the density matrix of $Q$, 
\begin{equation} 
\rho \rightarrow \frac{\mathcal{M}_{s}(\rho) }{P(s|\rho)} . \label{NORME} 
\end{equation}

\begin{figure}[t]
\includegraphics[width=\columnwidth]{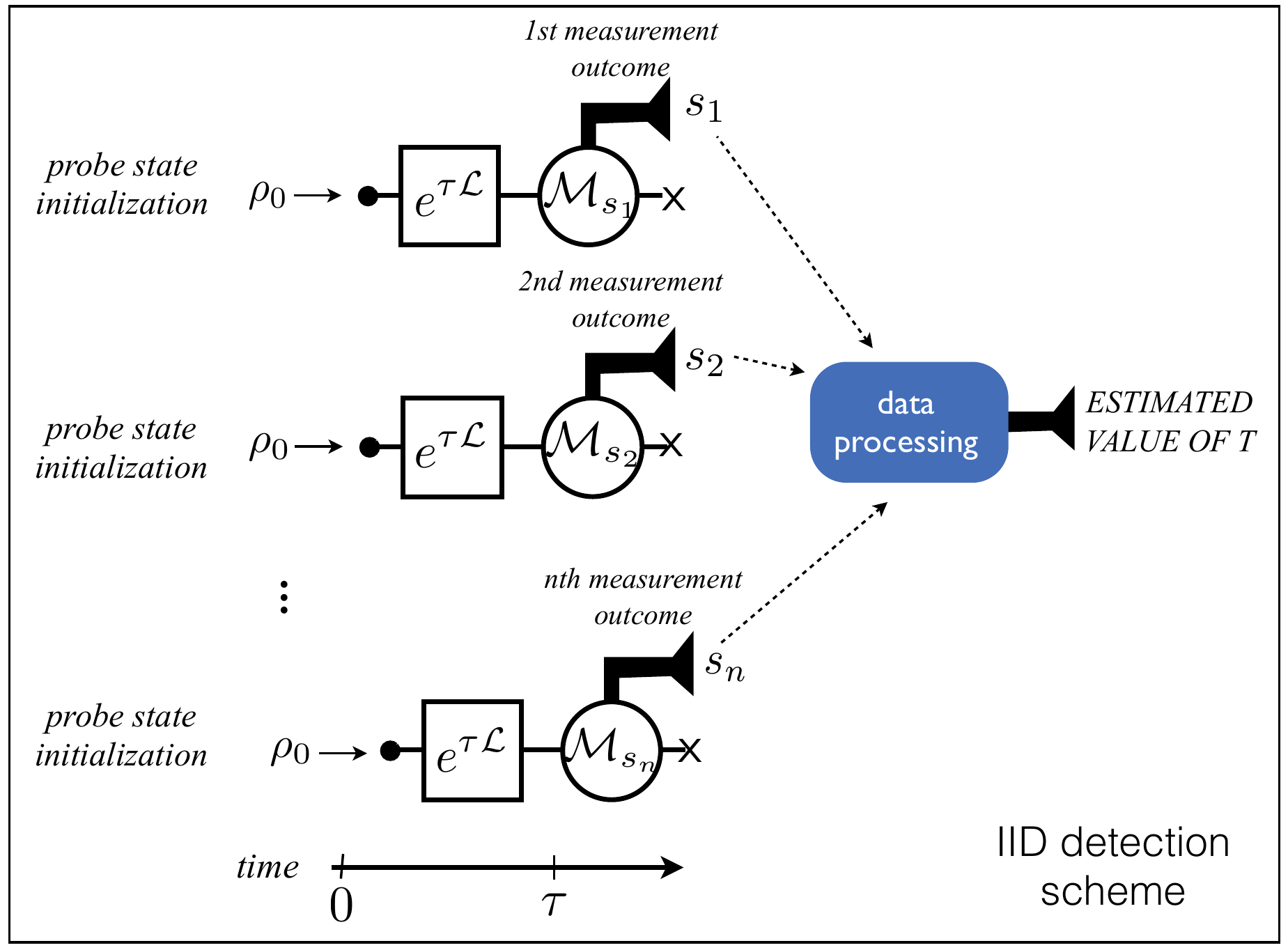} 
\caption{
Schematic representation of an estimation strategy of the bath temperature based on an IID detection scheme. Thermalization processes are represented by  the rectangular elements in the figure while  the measurement stages by the circular ones.  The time intervals of the thermalization events described by the maps $e^{\tau\mathcal{L}}$ are assumed to be uniform and identical to $\tau$, while the measurement events described by the maps $\mathcal{M}_s$ are assumed to be instantaneous. After each detection the  probe state is discarded and reinitialized into the input state $\rho_0$. The blue box represents a classical data processing which aims to produce an estimation of the temperature $T$ from the measurement outcomes $\{s_1, s_2, \ldots, s_n\}$.}\label{fig:iid}
\end{figure}

We then analyze two alternative scenarios. 
The first one corresponds to organizing $n$ detection events associated with the selected POVM in  an IID detection scheme as  shown in Fig.~\ref{fig:iid}. Here one tries to recover the bath temperature $T$ by repeating $n$ times the same experiment consisting in three basic steps, within the hypothesis \textit{iii)}: 
\begin{itemize} 
\item[{\it 1)}] initialization of the probing system  $Q$ in a selected input state $\rho_0$;
\item[{\it 2)}] evolution of $Q$ according to \eqref{LE} for a time interval $\tau$,  obtaining
\begin{equation} \label{RHOTAU} 
\rho(\tau) = e^{\tau \mathcal{L}}(\rho_0);
\end{equation} 
\item[{\it 3)}]  measurement of the selected POVM on the outcome state of the probe~(\ref{RHOTAU}).
\end{itemize} 
As a result one obtains an $n$-long sequence of outcomes  $\vec{s}=(s_1, s_2, \ldots, s_n)$  distributed according to the probability
\begin{equation} \label{NEWPROB}
 P^{(n)}_{\mathrm{IID}}(\vec{s}\,|\rho_0;\tau)= \prod_{j=1}^n P(s_j |\rho(\tau)),
 \end{equation} 
with $P(s_j |\rho(\tau))$ as in (\ref{EXPP10}). 
%%%%%%
%%%%%%
\begin{figure}[t]
\includegraphics[width=\columnwidth]{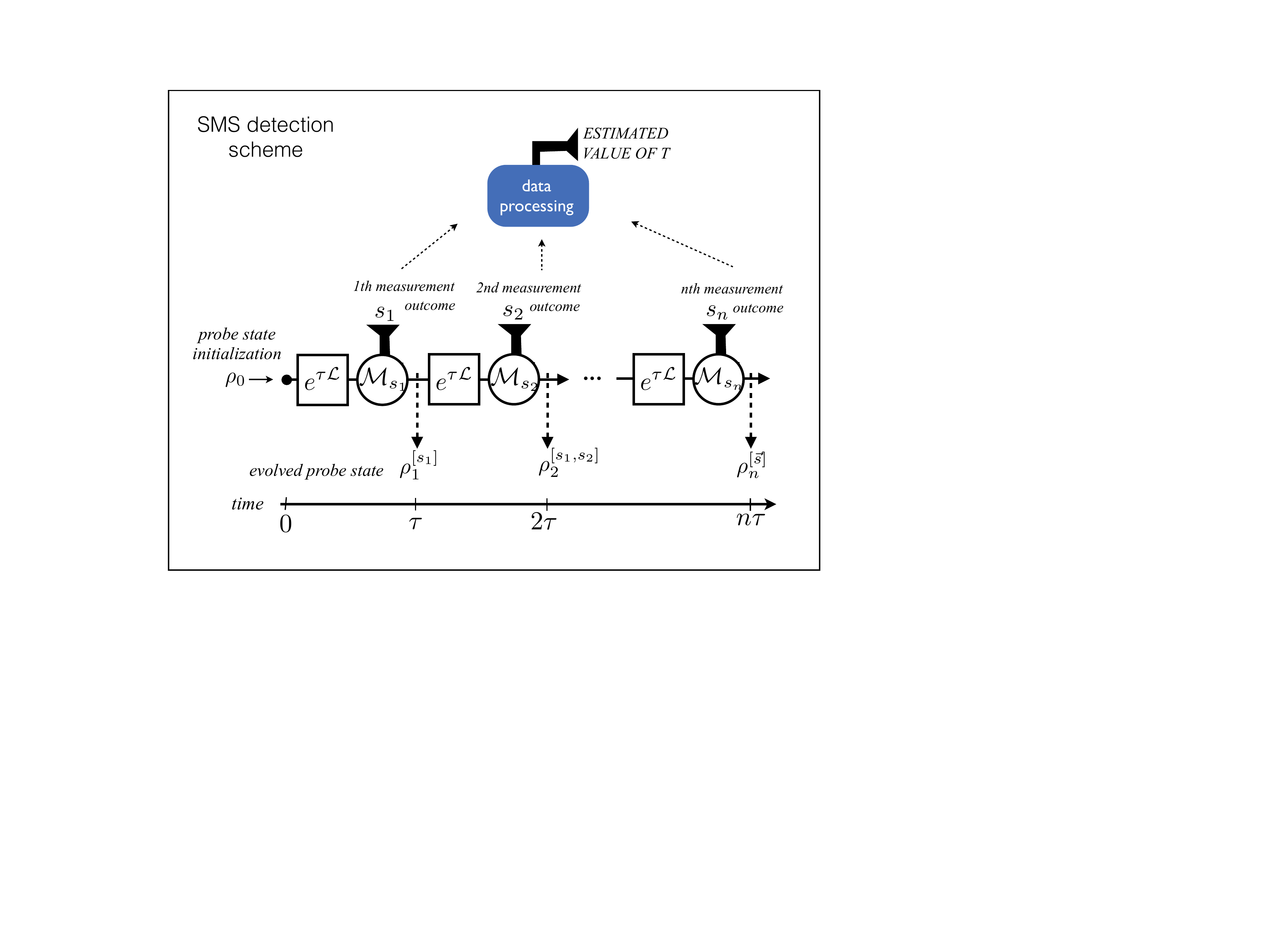} 
\caption{
Schematic representation of an estimation strategy of the bath temperature based on an SMS detection scheme. At variance with the IID scheme in Fig.~\ref{fig:iid} the state of the probe after the detection is not discarded or reinitialized. Instead it undergoes a series of subsequent thermalization events (rectangular elements in the figure) followed by measurements (circular elements). The time intervals of the thermalization events described by the maps $e^{\tau\mathcal{L}}$ are assumed to be uniform and identical to $\tau$, while the measurement events described by the maps $\mathcal{M}_s$ are assumed to be instantaneous. As in Fig.~\ref{fig:iid} the blue box represents a classical data processing on the acquired measurement data $\{s_1,s_2,\ldots,s_n\}$ to estimate the temperature $T$ of the bath.}\label{fig:seq}
\end{figure}
%%%%%%

The second scenario we consider is the SMS scheme of Fig.~\ref{fig:seq}, where 
after having  initialized $Q$ into the input state $\rho_0$, we let $Q$ interact with $B$ while we perform $n$ measurements described by a family of completely positive maps $\{\mathcal{M}_s\}$ at regular time intervals $\tau$ without re-preparing $Q$ after each measurement. Accordingly, indicating with $s_j$ the $j$th measurement outcome, every $\tau$ the system 
evolves one step ahead along the sequence of density matrices  $\rho_0$, $\rho^{[s_1]}_1$, $\rho^{[s_1,s_2]}_2$, \ldots, $\rho^{[s_1,\ldots, s_{n}]}_n$ generated by the stochastic process~(\ref{NORME}), i.e.,
\begin{align}
\rho^{[s_1]}_1&= \frac{(\mathcal{M}_{s_1}\circ  e^{\tau \mathcal{L}})(\rho_0) }{\Tr[(\mathcal{M}_{s_1}\circ  e^{\tau \mathcal{L}})(\rho_0) ]}  &&(t=\tau)  ,\nonumber\\
\rho^{[s_1,s_2]}_2&= \frac{(\mathcal{M}_{s_2}\circ  e^{\tau \mathcal{L}})(\rho^{[s_1]}_1)}{\Tr[(\mathcal{M}_{s_2}\circ  e^{\tau \mathcal{L}})(\rho^{[s_1]}_1)]}  &&(t=2\tau), \nonumber\\
&\qquad\qquad\vdots&&\nonumber\\ 
\rho^{[s_1,\ldots, s_{n}]}_n&= \frac{(\mathcal{M}_{s_n}\circ  e^{\tau \mathcal{L}})(\rho^{[s_1,\ldots, s_{n-1}]}_{n-1})}{\Tr[(\mathcal{M}_{s_n}\circ  e^{\tau \mathcal{L}})(\rho^{[s_1,\ldots, s_{n-1}]}_{n-1})]} && (t=n\tau),
\end{align}
 where the symbol ``$\circ$'' represents the  composition of superoperators and where we assume to neglect the time required by each measurement process.
Each of the normalization coefficients entering the above expressions gives the probability that a certain outcome takes place at each measurement step.  Therefore the probability that after $n$ steps
we obtain a certain string  $\vec{s}=(s_1, s_2, \ldots, s_n )$ of results is given by the product of the normalization coefficients
\begin{align}
P^{(n)}_{\mathrm{SMS}}(\vec{s}\,|\rho_0;\tau)
&=\prod_{j=1}^n P(s_j | \rho^{[s_1,\ldots, s_{j-1}]}_{j-1}(\tau))
\nonumber\\
&= \prod_{j=1}^n \Tr[( \mathcal{M}_{s_j} \circ e^{\tau \mathcal{L}} )
 (\rho^{[s_1,\ldots, s_{j-1}]}_{j-1})]  ,
\label{eq:p_seq1}
 \end{align} 
or equivalently,
\begin{multline}
P^{(n)}_{\mathrm{SMS}}(\vec{s}\,|\rho_0;\tau)
= \Tr[(\mathcal{M}_{s_n} \circ e^{\tau \mathcal{L}} \circ \mathcal{M}_{s_{n-1}}  \circ e^{\tau \mathcal{L}} \\
 {}\circ\cdots \circ \mathcal{M}_{s_1} \circ e^{\tau \mathcal{L}}) (\rho_0)] , \label{eq:p_seq}
\end{multline}
with $\rho^{[s_1,\ldots, s_{j-1}]}_{j-1}(\tau) = e^{\tau \mathcal{L}} ( \rho^{[s_1,\ldots, s_{j-1}]}_{j-1})$ being the evolution from $\rho^{[s_1,\ldots, s_{j-1}]}_{j-1}$ under the action of the 
process~(\ref{LE}) for time $\tau$.

In both IID and SMS scenarios detailed above, all the information one can recover about $T$ is stored in the statistical distribution of the associated outcome strings $\vec{s}$. 
A fair comparison between these two detection strategies can hence be obtained by invoking  the Cram\'er-Rao bound~\cite{cramer,paris1,metrology1,paris2,metrology2}. It establishes
that 
the minimum value of the root mean square error (RMSE)  $\Delta T^{(n)}$  one can get when trying to estimate the parameter $T$ from a sequence of $n$ outcomes $\vec{s}$ distributed according to the
probability $P^{(n)}(\vec{s})$ is given 
 by the inverse square root of the associated FI
\begin{equation}\label{eq:F_nNEW}
  \mathcal{F}^{(n)}=   \sum_{\vec{s}} \frac{1}{P^{(n)}(\vec{s})} \left(\frac{\partial  P^{(n)}(\vec{s})}{\partial T}\right)^2 ,
\end{equation}
i.e.,
  \begin{equation}\label{CR}
\Delta T^{(n)} \geq \frac{1}{\sqrt{  \mathcal{F}^{(n)}}} .
\end{equation} 
Accordingly larger values of $\mathcal{F}^{(n)}$ indicate the possibility of reaching higher levels of estimation accuracy. 
For the IID strategy this implies 
  \begin{equation}\label{CRIID}
\Delta T^{(n)}_{\mathrm{IID}}\geq \frac{1}{\sqrt{  \mathcal{F}^{(n)}_{\mathrm{IID}}(\rho_0;\tau)}} =  \frac{1}{ \sqrt{ n\mathcal{F}(\rho(\tau))}} , 
\end{equation} 
the $n^{-1/2}$ scaling being the trade-mark of the IID procedure. Indeed the following identity holds:
\begin{align}\label{eq:F_nIID}
  \mathcal{F}^{(n)}_{\mathrm{IID}}(\rho_0;\tau) &= \sum_{\vec{s}} \frac{1}{P^{(n)}_{\mathrm{IID}}(\vec{s}\,|\rho_0;\tau)} \left(\frac{\partial  P^{(n)}_{\mathrm{IID}}(\vec{s}\,|\rho_0;\tau)}{\partial T}\right)^2  \nonumber \\
  &= n   \mathcal{F}(\rho(\tau)), 
\end{align}
where    
\begin{equation} \label{FIS1} 
\mathcal{F}(\rho(\tau))= \sum_{s}  \frac{1}{P({s}|\rho(\tau))} \left(\frac{\partial  P({s}|\rho(\tau))}{\partial T}\right)^2 
\end{equation}
is the FI of  the probability~(\ref{EXPP10}) associated to the state (\ref{RHOTAU}). 
For the SMS strategy, instead, 
 the  Cram\'er-Rao bound yields
 \begin{equation} \label{CRSMS}
\Delta T^{(n)}_{\mathrm{SMS}}\geq \frac{1}{\sqrt{  \mathcal{F}^{(n)}_{\mathrm{SMS}}(\rho_0;\tau)}}  , 
\end{equation} 
with
\begin{equation}\label{eq:F_nSMS}
\mathcal{F}^{(n)}_{\mathrm{SMS}}(\rho_0;\tau) =   \sum_{\vec{s}} \frac{1}{P^{(n)}_{\mathrm{SMS}}(\vec{s}\,|\rho_0;\tau)} \left(\frac{\partial  P^{(n)}_{\mathrm{SMS}}(\vec{s}\,|\rho_0;\tau)}{\partial T}\right)^2
\end{equation}
being the FI associated to  the probability~(\ref{eq:p_seq}), which, at variance with (\ref{eq:F_nIID}), in general does not exhibit the same linear scaling with respect to $n$.

In the next section we shall study the behavior of (\ref{eq:F_nIID}) and (\ref{eq:F_nSMS}) for fixed choices of the POVM operators $\mathcal{M}_s$ and for fixed values of the number of the iterations $n$. In particular, we shall 
focus on the functional dependence of these quantities with respect to the input state $\rho_0$ of the probe.

%%%%%%%%%%%%%%%
\section{Comparing IID and SMS Strategies}\label{sec:comp} 
In what follows we consider
detection procedures which try to recover $T$ by  monitoring, with a certain accuracy, the populations of the energy levels of the probe. 
To describe the measurement process we select the following family of completely positive quantum maps 
 \begin{equation}\label{def:measure}
\mathcal{M}_s ({\cdots}) = M_s ({\cdots}) M_s^\dagger \quad (s=\pm), 
\end{equation}
with 
\begin{align}
M_{+} &=\Pi_{+}\cos\varphi + \Pi_-\sin\varphi,\nonumber\\
M_{-} &=\Pi_{+}\sin\varphi + \Pi_-\cos\varphi,\label{def:measure1}
\end{align}
where $\Pi_{\pm}= (\openone \pm \sigma_z)/2$  are  the projectors on the  eigenvectors of the qubit Hamiltonian $\frac{1}{2}\hbar\Omega\sigma_z$ of the probe $Q$ and where 
  the parameter $\varphi\in [0,\pi/4]$ gauges the effectiveness  of the  measurement as well as the
 disturbance it induces on $Q$.  Specifically, 
  for $\varphi=0$ the selected  POVM corresponds  to a projective measurement which induces stochastic jumps~(\ref{NORME}) into the probe energy eigenstates. As $\varphi$ increases,  the sharpness of the detection decreases to the extent  that  
for  $\varphi=\pi/4$ no information on $Q$ is gathered and the transformations (\ref{NORME}) results into the identity mapping.

 With this choice from (\ref{EXPP10}) and  (\ref{RHOTAU}) we obtain 
\begin{equation} 
  P(s|\rho(\tau))= (1 + s  \langle \sigma_z(\tau) \rangle \cos2\varphi )/2, 
\end{equation} 
 so that (\ref{FIS1}) becomes 
 \begin{equation}
\mathcal{F}(\rho(\tau))\label{REP}
=\frac{\cos^22 \varphi}{1-\langle \sigma_z (\tau)  \rangle^2\cos^22 \varphi } \left(\frac{\partial \langle \sigma_z (\tau)  \rangle}{\partial T}\right)^2, 
\end{equation}
where for a generic $t\geq 0$, $\langle \Theta (t) \rangle$ is a shorthand notation to represent the expectation value of the operator $\Theta$ on the evolved state of the probe at time $t$ under the thermalization map, i.e.,
\begin{equation} \langle \Theta (t) \rangle :=  \Tr[ \Theta \rho(t)] =\Tr[ \Theta e^{t \mathcal{L}} (\rho_0 ) ] . 
\end{equation} 
Explicit values of the above quantities can be obtained by direct integration of the equation of motion~(\ref{LE}),  which implies
\begin{equation}
 \langle \sigma_z (\tau) \rangle= e^{- \gamma \tau  \coth\frac{\beta}{2}}  \langle\sigma_z (0) \rangle - (1- e^{- \gamma \tau   \coth\frac{\beta}{2}}) \tanh\frac{\beta}{2}.  \label{eq:components}
\end{equation}
 Similarly, from (\ref{eq:p_seq1}) we get  
  \begin{equation} \label{PSMS} 
P^{(n)}_{\mathrm{SMS}}(\vec{s}\,|\rho_0;\tau)=\frac{1}{2^n} \prod_{j=1}^n(1 + s_j  \langle \sigma_z (\tau)\rangle_{j} \cos2\varphi ), 
\end{equation}
where  for $j=1,\ldots, n$ we define 
\begin{equation} 
\langle \sigma_z (\tau)\rangle_{j}  : = \Tr [\sigma_z e^{\tau \mathcal{L}}(\rho^{[s_1,\ldots, s_{j-1}]}_{j-1})] ,\end{equation} 
which reads as in (\ref{eq:components}) with $\langle\sigma_z (0) \rangle$ replaced by $\Tr [\sigma_z \rho^{[s_1,\ldots, s_{j-1}]}_{j-1}]$.

%%%%%%%%%%%
Replacing (\ref{REP}) into (\ref{eq:F_nIID}), and (\ref{PSMS}) into (\ref{eq:F_nSMS}) we can now study the FI of the two procedures 
for different choices of the POVM parameter $\varphi$ and for different values of the iterations $n$. 
In particular in the following subsections we shall focus on the dependence   upon the input state $\rho_0$
 of the probe $Q$. 
For the sake of simplicity and without loss of generality,  the times $t$ (or $\tau$) will be parametrized in units of the coupling constant $\gamma$, and the bath temperature $T$ in units of the qubit energy gap $\hbar \Omega / k_B$.
We anticipate that for both SMS and IID cases, the FI exhibits a functional dependence upon $T$, which presents a single peak and vanishes in the limit of zero and infinite temperature. 
On one hand, these last facts can be justified by reminding that, as evident from (\ref{eq:F_nNEW}), FI is an increasing functional of the first derivative in $T$ of the  probability distribution $P^{(n)}(\vec{s})$. Accordingly it  accounts for the sensitivity of the probing system $Q$ under small variations of the bath temperature (see also Ref.~\cite{betaloc}). In other words, the more the final state of the qubit is affected by slight variations of the bath temperature $T$, the higher are the values of  the associated FI\@. 
At zero temperature, the bosonic bath $B$ is frozen in its ground state, a situation which is almost unaltered even if one increases $T$ by an infinitesimal amount $\delta T$. 
In this case $Q$ is almost insensitive to the small variations in $T$, yielding a vanishingly small value of FI\@. 
An analogous scenario emerges in the opposite limit of infinite temperature, where all the energy levels of the bath $B$  are equally populated, a situation which for all practical purposes is basically unaltered by infinitesimal changes of the order of $\delta T$. On the other hand, the presence of a single maximum in the $T$ dependence of the FI can be finally linked to the 
structure of the thermometer spectrum characterized by a single energy gap, a fact which was
also pointed out in Ref.~\cite{correa1} for the specific case of projective measurements ($\varphi=0$) and $\rho_0=\Pi_-$.
\begin{figure}[t!]
\includegraphics[width=\columnwidth]{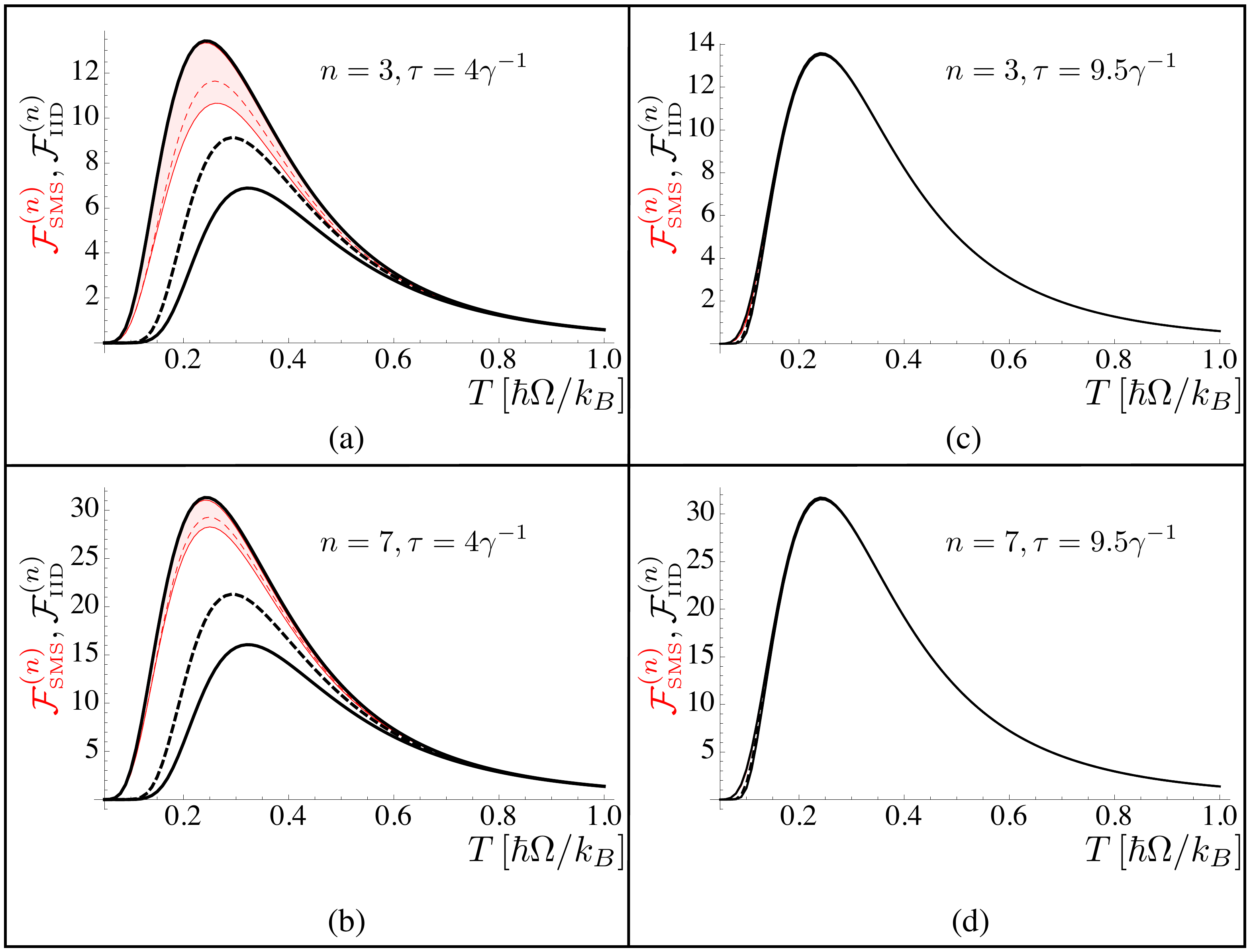} 
\caption{
Plots of the FI associated with the IID and SMS procedures where the POVM  defined in (\ref{def:measure}) and (\ref{def:measure1}) represents a projective measurement ($\varphi=0$) on the probe qubit $Q$. The panels refer to different values of the number of repetition $n$ and of the interaction time $\tau$. Thick (black) lines correspond to the IID protocol while the thin (red) ones refer to the SMS (the region corresponding to the SMS protocol is shaded).  In both cases, the uppermost and lowest solid lines refer to the optimal and worst choices of the input state $\rho_0$, while the dashed lines in between refer to the average of the FIs over a uniform (Haar) sampling of the blanced purification of the input state (see the main text). Notice that for long interaction times [panels (c) and (d)] all the curves collapse to a single one.} \label{fig:proj}
\end{figure}

%%%%%%%%%%%%%%
\subsection{Projective measurements, $\bm{\varphi=0}$}\label{varphi0}
We start by considering the case in which the measurement operators (\ref{def:measure1}) reduce to rank-one projectors on the ground and excited energy levels, $M_- = \Pi_-$ and $M_+=\Pi_+$, respectively.
In panels (a) and (b) of Fig.~\ref{fig:proj} we set $\tau=4 \gamma^{-1}$ and plot $\mathcal{F}^{(n)}_\mathrm{IID}(\rho_0;\tau)$ and $\mathcal{F}^{(n)}_\mathrm{SMS}(\rho_0;\tau)$ for $n=3$ and $n=7$, respectively. 
We consider a uniform sampling of the input probe state $\rho_0$, induced by the Haar measure over the unitary group acting on the Hilbert space associated to the balanced purification \cite{mixmatrix}: that is, we sample pure states $\ket{\Psi_0}$ uniformly on an extended Hilbert space $\mathcal{H}_Q\otimes\mathcal{H}_{Q'}=\mathbb{C}^2\otimes\mathbb{C}^2$, on which any mixed states of the qubit probe $Q$ can be purified, to get input probe states through $\rho_0=\Tr_{Q'}\ket{\Psi_0}\bra{\Psi_0}$, i.e.~through partial trace over the auxiliary Hilbert space $\mathcal{H}_{Q'}$.
From the Cram\'er-Rao bound \eqref{CR} [and therefore from \eqref{CRIID} and \eqref{CRSMS}] it follows that the uppermost and lowest solid lines refer to the optimal and worst choices of the input state $\rho_0$, which in both schemes we have numerically proved to coincide with the ground state and with the first excited level, respectively. In between, the dashed lines refer to the average values of the FI over the sampled input probe states $\rho_0$.
\begin{figure}[t!]
\includegraphics[width=0.8\columnwidth]{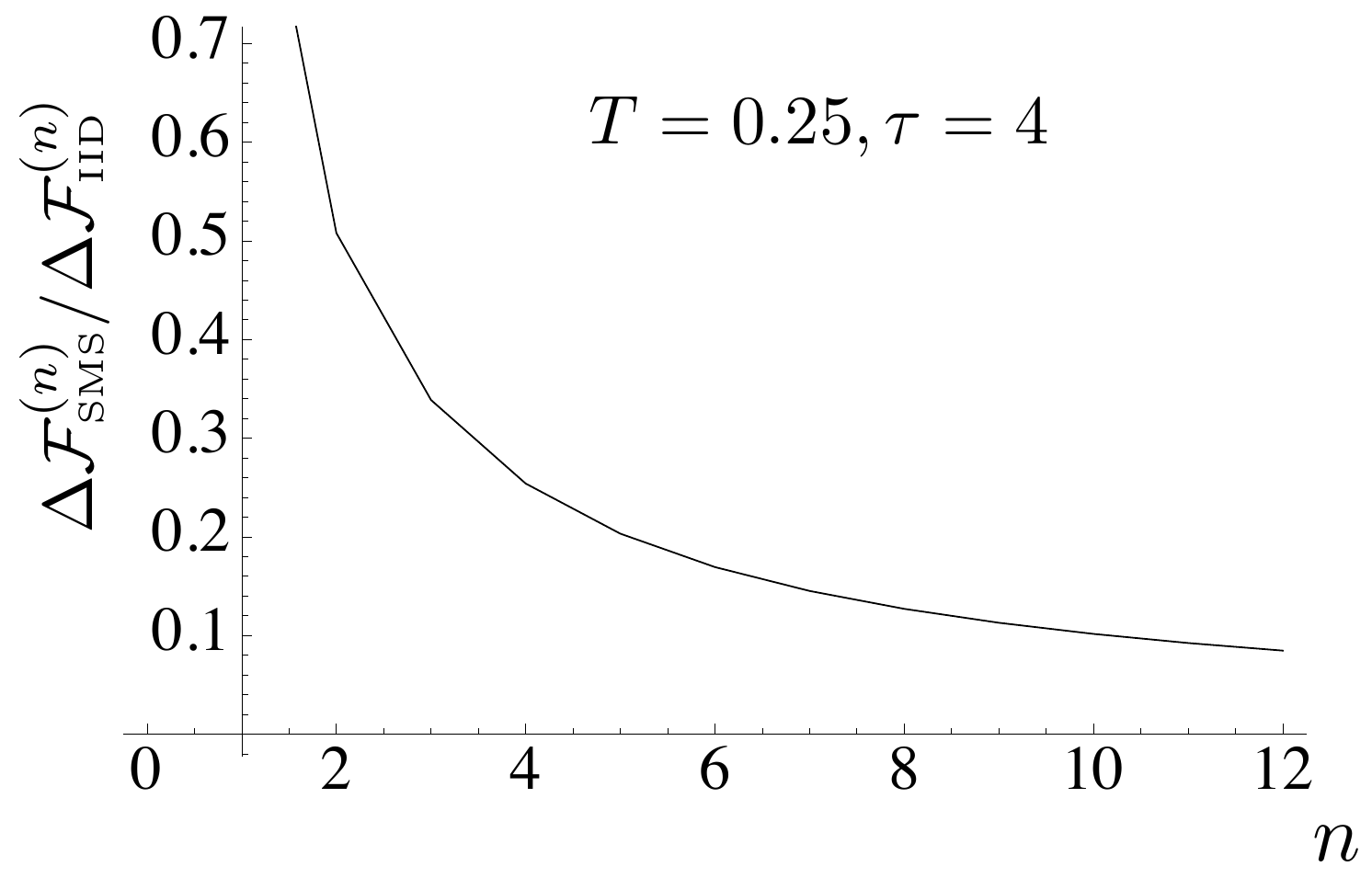}
\caption{$\Delta\mathcal{F}^{(n)}_\mathrm{SMS}$ and $\Delta\mathcal{F}^{(n)}_\mathrm{IID}$ are the max-min band widths of the FI associated to the choice of the input states in the SMS and IID schemes, respectively. Here we plot their ratio as a function of the number of measurements $n$ performed on the probe. It results that $\Delta\mathcal{F}^{(n)}_\mathrm{SMS}$ shrinks  more rapidly than its counterpart for the IID scheme.}\label{fig:fmax2}
\end{figure} 
If we initialize $Q$ in the ground state $\rho_0=\Pi_-$ (upper solid lines),  $\mathcal{F}^{(n)}_\mathrm{IID}(\rho_0;\tau)$  coincides with the so-called quantum Fisher information (QFI), giving the highest achievable accuracy for the bath temperature reconstruction through a qubit probe~\cite{burgarth}.
 With the same choice of $\rho_0$ also $\mathcal{F}^{(n)}_\mathrm{SMS}(\rho_0;\tau)$ gets its maximum value, but the IID strategy always slightly outperform the SMS strategy, i.e.~$\mathcal{F}^{(n)}_\mathrm{IID}(\rho_0;\tau) > \mathcal{F}^{(n)}_\mathrm{SMS}(\rho_0;\tau)$.
For non-optimal input states an interesting phenomenon is observed: for all bath temperatures the SMS protocol, both on average and in the worst case scenario, offers a better performance with respect to the IID protocol. 
Notice also that the gap between the FIs by the optimal and worst choices of the input state shrinks with $n$ more rapidly in the SMS protocol than in the IID scheme: see Fig.~\ref{fig:fmax2}. Stated differently, the SMS protocol is less affected by the choice of the input probe state, thus providing  a higher versatility with respect to the standard IID  measurement scheme. 
However, for sufficiently long interaction times between the measurements, $\tau \gg \gamma^{-1}$, the four curves collapse, thus giving the same accuracy irresspective of the chosen protocol and input probe state: see Figs.~\ref{fig:proj}(c) and (d). This is a consequence of the fact that in this regime the mapping (\ref{RHOTAU}) ensures
complete thermalization of $Q$  by bringing it 
 into the  fixed point $\rho_\mathrm{th}=e^{-\beta\sigma_z} /\Tr[e^{-\beta \sigma_z}]$  independently of the input state.

%%%%%%%%%%%
\subsection{Non-projective measurements, $\bm{\varphi\neq0}$}
\begin{figure}[t!] 
\includegraphics[width=1\columnwidth]{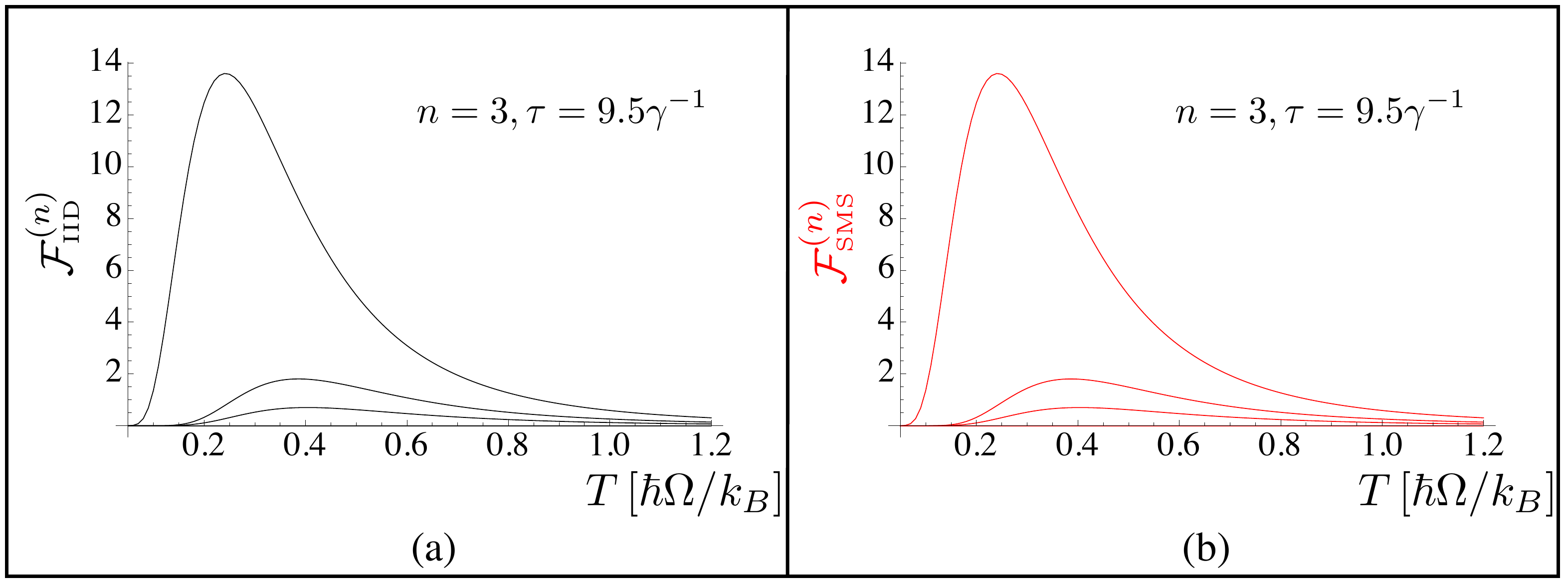}  
\caption{
Plots of the FI for the IID (a) and SMS (b) procedures by setting $n=3$, $\rho_0=\Pi_-$, and $\tau=9.5\gamma^{-1}$. 
The three curves in each panel correspond to the FIs by different strengths of the measurements, setting in (\ref{def:measure}) and (\ref{def:measure1}) for the POVM $\varphi=0,\pi/8,\pi/6$, from the uppermost curve to the lowermost curve. 
The case $\varphi=\pi/4$ corresponds to the random completely uninformative measurement and yields $\mathcal{F}^{(n)}_\mathrm{IID}(\rho_0;\tau)=\mathcal{F}^{(n)}_\mathrm{SMS}(\rho_0;\tau) = 0$, for all $n$ and $T$.} \label{fig:proj_nonproj}
\end{figure}

\begin{figure}[t!] 
\includegraphics[width=1\columnwidth]{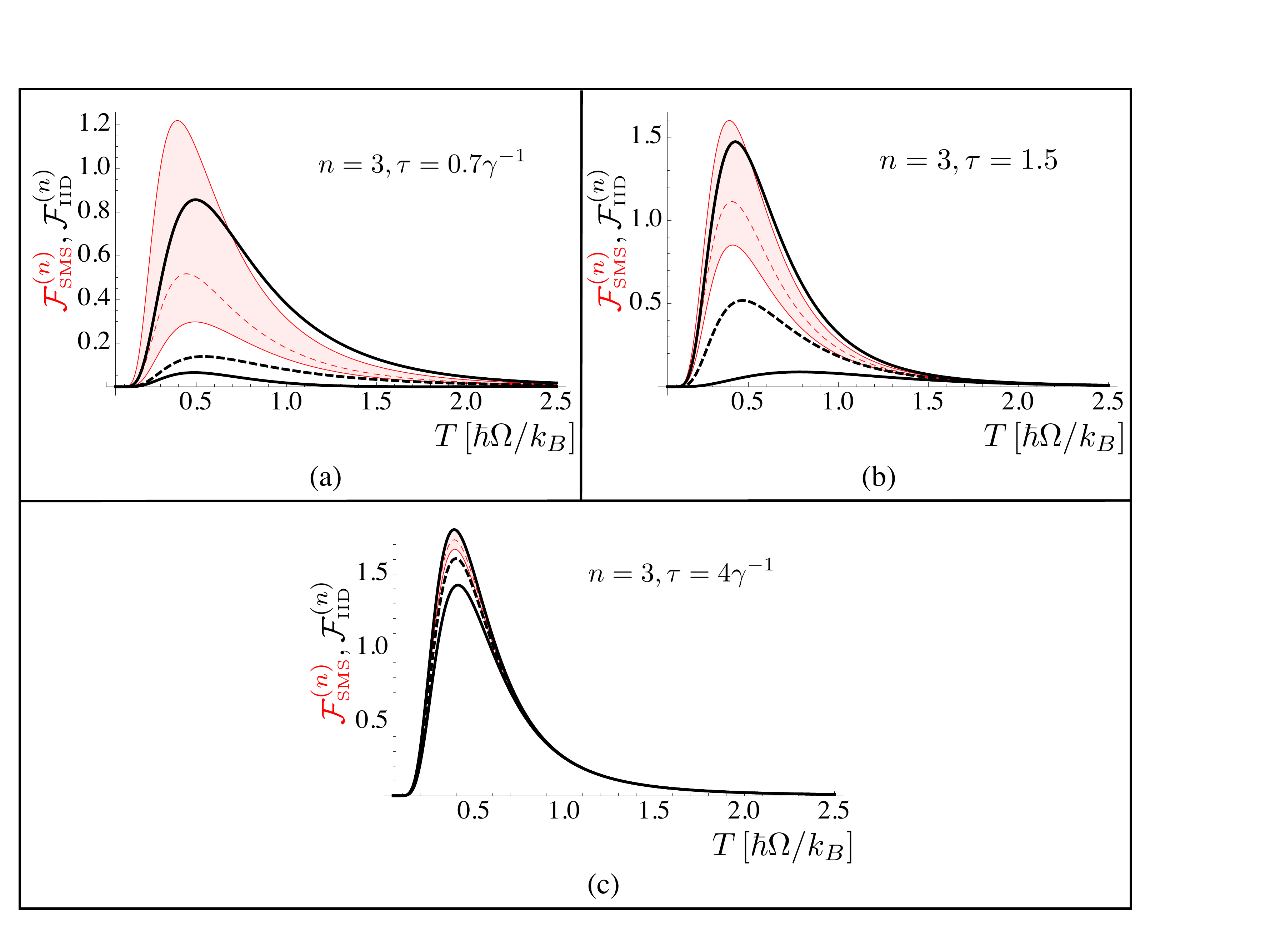}
\caption{Plots of the FI associated with the IID and SMS procedures for the case in which the POVM defined in (\ref{def:measure}) and (\ref{def:measure1}) represents a non-projective measurement ($\varphi=\pi/8$) on the probe qubit $Q$. We have used the same convention as in Fig.~\ref{fig:proj} in order to distinguish the curves corresponding to the IID protocol and SMS procedures, computed for different values of the interaction time $\tau$ for the number of repetition $n$. Notice that for sufficiently small interaction times the SMS performs better than the IID scheme, for certain values of $T$, even for the optimal input states.} \label{fig:non_proj}
\end{figure}

When setting $\varphi \neq 0$ the POVM~(\ref{def:measure1}) describes  non-projective noisy measurements,  which are less informative but less disruptive on the state of the probe $Q$. As evident from Fig.~\ref{fig:proj_nonproj}, for fixed choices of $\rho_0$, $n$, and $\tau$, the  accuracy of the procedure degrades as
$\varphi$ increases, both in the IID and in the SMS scenario. 
%%%%%%%
%%%%%%%%
In Fig.~\ref{fig:non_proj}  we present  instead a comparison between the sensitivities of the two approaches with respect to the choice of the input state $\rho_0$ focusing on the 
case of $\varphi=\pi/8$. 
Differently from the projective measurement case discussed in Sec.~\ref{varphi0}, we have that if the interaction time between the thermometer and the bath is sufficiently small, $\tau \lesssim 2.5\gamma^{-1}$, the SMS performs better than the IID scheme (at least for certain values of the bath temperature $T$), even for the optimal input states. 
See panels (a) and (b) of Fig.~\ref{fig:non_proj}. 
Once more we interpret  this result as a consequence  of the fact that, at variance with the IID scheme, in an SMS procedure the probe is forced to adapt to the bath: in the case of the noisy POVM analyzed here and in the presence of a small interaction interval $\tau$,  this mechanism is powerful  enough  to give an advantage also in terms of the maximum sensitivity achievable. Besides this peculiar effect, we have that the SMS still proves to be also more versatile than the IID scheme  as it provides 
a better performance not only on average but also for the worst possible choice of the input state. 

\section{Conclusions}\label{sec:conclusions}
In this work we focused on the problem of determining the temperature of an external reservoir
via the measurements on a probe $Q$ that is put in thermal contact with the reservoir, and plays the role of thermometer. 
In this framework we compare the performances of two alternative scenarios: the IID scheme, where 
$Q$ is measured and re-prepared a certain number of times, and the SMS~\cite{burgarth, guta, catana, kiilerich} where instead the same  detections are performed but in sequence without  intermediate state reinitializations. 
The aim of the analysis is to study the dependence of these procedures with respect to the choice of the input state  $\rho_0$ of $Q$. 
Our findings, while deriving from a specific model (i.e.~a qubit probe in thermal contact with a Bosonic reservoir monitored via
a noisy POVM which reads the populations of its energy levels) are indicative of the fact that the SMS approach is more versatile than the IID approach with respect to $\rho_0$. 
This can be associated with the fact that in the SMS $Q$ is slowly drifting toward a fixed point configuration independently of the input state we have selected~\cite{burgarth}.  Therefore at variance with the IID configuration, one expects that in the SMS approach there would be no  choices for  $\rho_0$ which are really  ``bad'': the reservoir will 
``guide'' any possible input towards a relatively good configuration, protecting hence the estimation procedure from unwanted initialization errors.

\acknowledgments
This work was supported by the EU Collaborative Project TherMiQ (Grant agreement 618074), by the EU project COST Action MP1209 Thermodynamics in the quantum regime, and by the Top Global University Project from the Ministry of Education, Culture, Sports, Science and Technology (MEXT), Japan.
KY was supported by the Grant-in-Aid for Scientific Research (C) (No.\ 26400406) from the Japan Society for the Promotion of Science (JSPS) and by the Waseda University Grant for Special Research Projects (No.\ 2016K-215).


\begin{thebibliography}{10}
\bibitem{NOTA}
Even though not explicitly related to the thermometric tasks, the idea of
  determining unknown coefficients of phenomenological master
  equations through the interaction of an external and controllable probe with
  the bath has already been explored, for example in
  the context of quantum symplectic tomography
  \cite{bellomo1,bellomo2,bellomo3,zang,thapliyala}.
\end{thebibliography}
\end{document}